\documentclass[12pt]{article}
 \newcommand{\be}[1]{\begin{equation}\label{#1}}
 \newcommand{\ba}[1]{\begin{eqnarray}\label{#1}}
 \newcommand{\ep}[1]{\epsilon_{#1}}

 \newcommand{\de}[1]{\delta_{#1}}

 \newcommand{\rd}{{\rm d}}

 \newcommand{\pa}[1]{\left(#1\right)}
 \newcommand{\paq}[1]{\left[#1\right]}

 \newcommand{\M}{{\rm M_{\rm P}}}

 \def\ee{\end{equation}}
 \def\ea{\end{eqnarray}}

\usepackage{amsmath}

\usepackage{graphicx,latexsym}
\usepackage{graphicx,epsf}
\usepackage{color}
\usepackage{epsfig}
\usepackage{amsmath}
\usepackage[T1]{fontenc}
\usepackage[utf8]{inputenc}
\usepackage{tikz}
\usepackage{amssymb}
\usepackage{authblk}
\begin{document}
\title{An Effective Theory for Higgs Inflation}
\author[1]{Alessandro Tronconi\thanks{Alessandro.Tronconi@bo.infn.it}}
\author[1]{Giovanni Venturi\thanks{Giovanni.Venturi@bo.infn.it}}
\affil[1]{Dipartimento di Fisica e Astronomia, Universit\'a di Bologna and INFN, Via Irnerio 46, 40126 Bologna,
Italy}

\date{}

\maketitle
\begin{abstract}
The generation of large curvature perturbations associated with the production of primordial black holes is studied in the context of a Higgs inflaton. To enable this amplification, we consider an inflationary model in which the tree-level action for gravity and the Standard Model Higgs is modified by quantum corrections, described by a series of higher-dimension operators. Finally within a minimal EFT framework, we present two viable models in which the spectrum of curvature perturbations generated by the Higgs field is consistent with CMB observations and can lead to the formation of primordial black holes in the asteroid mass range, potentially accounting for the entirety of dark matter.
\end{abstract}
\section{Introduction}
Well before its discovery at the LHC~\cite{ATLAS2012,CMS2012}, the Higgs scalar had already been considered a viable candidate for the inflaton~\cite{Bezrukov:2007ep}. In order for the Standard Model (SM) Higgs to reproduce inflationary observables correctly, a very large non-minimal coupling between the Higgs field and the gravitational sector must be taken into account. Higgs inflation is nowadays regarded as one of the most promising (or at least most intriguing) hypotheses for describing the very early Universe, as it essentially requires no new physics and is in very good agreement with observations~\cite{Planck:2018jri}. The presence of a non-minimal interaction between the Higgs and the gravitational sector can indeed be justified as a consequence of quantum gravitational corrections, as originally proposed by Sakharov and Zee in their seminal works about Induced Gravity \cite{Sakharov:1967pk,Zee:1978wi}. Similar ideas also led to $R^2$ Starobinsky inflation \cite{Starobinsky:1980te} which, like Higgs inflation, fits inflationary observables very well. In fact, it has been realised that Induced Gravity and $R^2$ inflation are mathematically equivalent models that can be mapped into one another via a redefinition of the degrees of freedom.

Such models are essentially derived by attempting to include quantum corrections to matter and gravity. In their original formulation, their (effective) action contains first-order (one-loop) quantum effects from what is, in principle, an infinite tower of higher-order effective contributions, which become increasingly relevant at higher energies. From this perspective, classical General Relativity emerges as the low-energy limit of a more fundamental theory which, up to certain scales, can be formulated as an effective quantum field theory (EFT). At much higher, trans-Planckian, energies, one expects this effective description to break down, requiring a new (yet unknown) physics framework.

Over the past 45 years, a huge variety of inflationary models has been proposed. Many have already been ruled out by increasingly precise CMB observations, yet several remain compatible with current data. For these surviving models, we can constrain only a small portion---just a few e-folds---of the inflationary dynamics, which is imprinted in the CMB anisotropy spectrum at scales comparable to the size of the present Universe.

Understanding dark matter (DM) remains one of the greatest challenges in modern physics. It was proposed several years ago that DM could be entirely composed of primordial black holes (PBHs)~\cite{Novikov:1967tw, Hawking:1971ei}. These PBHs could have formed after the Hot Big Bang from the collapse of sufficiently large primordial density perturbations. They are ``cheap'' DM candidates, as they require no extensions to the Standard Model. If these over-densities are generated during inflation, then the inflationary dynamics can be constrained at energy scales far beyond the reach of the CMB---i.e. typically at much smaller scales.

The aim of this article is to investigate the production of large over-densities within the context of Higgs inflation. To this end, we consider an effective action for the SM Higgs coupled to gravity, incorporating a series of higher-order operators typically arising from quantum corrections. For simplicity, we assume a $\mathbb{Z}_2$ symmetry for the Higgs field, which restricts the number of allowed operators. In its original formulation, Higgs inflation contains only one free parameter---the non-minimal coupling $\xi$---and, quite remarkably, it fits CMB observables simply by appropriately tuning $\xi$. When quantum corrections are included, the number of free parameters increases; however, the requirement of generating large scalar fluctuations at some point during inflation imposes strong constraints on the possible values of such parameters, making the search for viable models highly non-trivial.\\
The article is organised as follows. In Sec. II, the general model is presented and the homogeneous dynamical equations are derived. In Sec. III, we develop the slow-roll (SR) approximation and obtain expressions for the inflationary observables associated with CMB anisotropies. In Sec. IV, we discuss the amplification of the curvature perturbation spectrum, and in Sec. V, we present two viable models of Higgs inflation. Finally, in Sec. VI, we draw our conclusions.

\section{The model}
Let us consider the following generalisation of the original Higgs inflation model: 
\begin{equation}
\label{action}
S=\int d^4 x \sqrt{-g}\left[ U(h)R-\frac{1}{2}g^{\mu\nu}h_{,\mu}h_{,\nu}+V(h)\right],
\end{equation}
where $g$ is the determinant of the metric tensor $g_{\mu\nu}$,
\be{Ugen}
U=\frac{\M^2}{2}\paq{1+\sum_{i} u_i \pa{\frac{h}{\M}}^{2i}}
\ee
and 
\be{V}
V=\frac{\lambda}{4}\pa{h^2-v^2}^2\paq{1+\sum_{i} v_i \pa{\frac{h}{\M}}^{2i}},
\ee
$v=246\,{\rm GeV}$ is the Higgs v.e.v. (and can be neglected during inflation when $h\gg v$) and $\lambda\sim 1/8$ is the Higgs self-coupling whose value is fixed by the observed value of the Higgs mass. On setting $v_i=0$, $u_1\equiv \xi$ and $u_{i>1}=0$ one recovers the vanilla model \cite{Bezrukov:2007ep}. \\
The evolution of a homogeneous Higgs field on the spatially
flat FRW universe with the metric 
\be{metric}
\rd s^2=\rd t^2-a^2(t)\rd\vec x^2,
\ee
is governed by the following Klein--Gordon equation
\begin{equation}
\label{dequsigma}
\pa{\ddot  h+3H\dot h}+V_{, h}=6\left(\dot H +2H^2\right)U_{, h}\,,
\end{equation}
where the subscript $ _{, h}$ indicates the derivative w.r.t. the Higgs field
$ h$.

The homogeneous Einstein equations are
\begin{equation}
\label{equ00}
6UH^2+6\dot U H=\frac{1}{2}{\dot{ h}}^2+V,
\end{equation}
\begin{equation}
\label{equ11}
2U\left(2\dot H+3H^2\right)+4\dot U H+2\ddot U +\frac{1}{2}{\dot{ h}}^2-V=0,
\end{equation}
and can be combined to obtain
\begin{equation}
\label{equ01}
4U\dot H-2HU_{, h}\dot{ h}+2\left(U_{, h h}{\dot{ h}}^2+U_{, h}\ddot{ h}\right) +{\dot h}^2=0.
\end{equation}
The slow-roll (SR) parameters hierarchy is used in the context of inflation as a replacement of the homogeneous degrees of freedom $H$ and $h$ (inflaton). The time variation of such parameters divided by $H$ is negligible if they are much smaller than unity.  Therefore, the adoption of the SR parameters instead of $H$, $h$ and their derivatives is useful to obtain an approximate form of the homogeneous dynamical equations. Moreover, the inflationary observables can be written in terms of the SR parameters. Throughout this article, we shall adopt two different hierarchies of the SR parameters. The Hubble flow function hierarchy is recursively defined as
\be{Hff}
\ep{0}=\frac{H_0}{H},\; \ep{i+1}=\frac{\dot{\ep{i}}}{H\ep{i}}\equiv \ep{i}^{-1}\frac{\rd{\ep{i}}}{\rd N}\Rightarrow \frac{\rd \ep{i}}{\rd N}=\ep{i}\ep{i+1}
\ee
and the scalar field flow function hierarchy is
\be{Sff}
\de{0}=\frac{h}{h_0}, \;\de{i+1}=\frac{\dot{\de{i}}}{H\de{i}}\equiv \de{i}^{-1}\frac{\rd{\de{i}}}{\rd N}\Rightarrow \frac{\rd \de{i}}{\rd N}=\de{i}\de{i+1}
\ee
where $H_0$ and $h_0$ are arbitrary constants and $N$ is the e-folding number defined by $\rd N=H\rd t$.
On substituting the SR parameters in (\ref{dequsigma}--\ref{equ11}) one finds the following homogeneous dynamical equations
\be{equ00sr}
H^2=\frac{V}{6 U\pa{1+n_U\de{1}-\frac{ h^2}{12U}\de{1}^2}},
\ee
\be{equ11sr}
\ep{1}=\frac{n_U\de{1}\pa{\de{2}+n_{n_U}\de{1}+n_U\de{1}-1}+\frac{ h^2}{2U}\de{1}^2}{2+n_U\de{1}},
\ee
\be{equationsr}
\frac{ h^2}{U}\de{1}\pa{\de{1}+\de{2}-\ep{1}+3}+\frac{V}{H^2U}n_V=6n_U\pa{2-\ep{1}}.
\ee
where the recurrent notation $n_f\equiv h\,f_{,h}/f$ indicates the logarithmic derivative of $\ln f(h)$.
Let us note that for polynomial $U$ where $h$ has, at most, powers $\mathcal{O}(10)$, $n_U$ and $n_{n_U}$ are usually $\mathcal{O}(1)-\mathcal{O}(10)$. In contrast $ h^2/U$ may be large. This, for example, occurs in Induced Gravity (see for example \cite{Cerioni:2009kn})with $U=\xi  h^2/2$ and $\xi\ll 1$. In such a case $n_U=2$, $n_{n_U}=0$ but $ h^2/U=\xi^{-1}\gg 1$. 
\section{Slow Roll regime}
The nearly scale invariant spectrum of the CMB photons can be explained by a slow rolling inflationary evolution i.e. by the assumption that the homogeneous degrees of freedom ``slowly vary'' during the early inflationary stages, namely $\sim 60$ e-folds before inflation ends. This slow variation is associated with SR parameters having values $\ll 1$. Formally, the SR approximation, consists in keeping contributions linear in the SR parameters in the dynamical equations and then neglecting higher order terms in such parameters. Since, by construction (see (\ref{Hff},\ref{Sff})), time derivatives of SR parameters are higher order terms, neglecting such contributions is equivalent to assuming that linear terms are constant.\\ 
In the general case we are discussing the SR approximation deserves care. In the exact equations (\ref{equ00sr}-\ref{equationsr}) the SR parameters are multiplied by functions of the scalar field which we generically indicate here as $n_f$. Such function are, for the cases of interest, rational functions taking values $\mathcal{O}(1)$. Thus, to the first order in the SR approximation, products of the form $n_f\de{i}$ can be considered constant since their derivative is of second order in the SR parameters: 
\be{srex1}
\frac{\rd \,n_f\de{i}}{\rd N}= n_f \de{i}\de{i+1}+n_{n_f} n_f \de{1}\de{i}\simeq \mathcal{O}(\de{i}^2).
\ee
We further observe that any dimensionless function of $h$, $g(h)$, when differentiated, leads to
\be{srex2}
\frac{\rd g(h)}{\rd N}=g \,n_g\, \de{1}.
\ee
This derivative is negligible if $g\ll 1$ and thus $g$ can be considered a constant. The expression $\de{1}^2 h^2/U$ with $ h^2/U\gg 1$ is essentially equivalent to $n_f\de{i}$ and can be treated similarly. One that has that (\ref{equ11sr}) can be generally approximated by 
\be{equ11srappIG}
\ep{1}\simeq-\frac{n_U}{2}\de{1}+\frac{h^2}{4U}\de{1}^2,
\ee
during SR. The last contribution is non-negligible when $h^2/U\gg 1$ and becomes negligible for $ h^2/U\le 1$.
On substituting (\ref{equ00sr}) and (\ref{equ11sr}) in (\ref{equationsr}) one then finds the following equation
\ba{sreq}
&&\frac{ h^2}{6U}\de{1}\paq{\de{1}+\de{2}+1}+n_V\pa{1+n_U\de{1}-\frac{ h^2}{12U}\de{1}^2}\nonumber\\
&&=\pa{n_U-\frac{ h^2}{6U}\de{1}}\paq{2-\frac{n_U\de{1}\pa{\de{2}+n_{n_U}\de{1}+n_U\de{1}-1}+\frac{ h^2}{2U}\de{1}^2}{2+n_U\de{1}}}
\ea
where the dependence on $H$ and its derivatives has been simplified. Since
\be{de2def}
\de{2}\equiv \de{1}^{-1}\frac{\rd \de{1}}{\rd N}=\de{1}^{-1}\frac{\rd  h}{\rd N}\frac{\rd \de{1}}{\rd  h}= h \frac{\rd \de{1}}{\rd  h}
\ee
if $h$ is a monotonic function of $N$, then (\ref{sreq}) is a second order differential equation for $ h(N)$ or a first order differential equation for $\de{1}( h)$. We observe that the de Sitter solution $\ep{i>0}=\de{i>0}=0$ exists if $n_V=2n_U$ i.e. $V\propto U^2$. Moreover in the SR limit $\de{i}\de{j}\ll \de{l}$ for arbitrary $i,j,l$, and Eq. (\ref{sreq}) takes the form
\be{sreqSR}
\de{1}\simeq \frac{2n_U-n_V}{\frac{ h^2}{2U}+n_U\pa{n_V-n_U/2}}.
\ee
We observe that $\de{1}$ does not depend on the value of the potential, which enters (\ref{sreqSR}) through $n_V$, but depends on the value of $U$.\\

Inflationary observables can be obtained on explicitly solving the Mukhanov-Sasaki equation \cite{Hwang:1995bv}
\be{MS}
v''_k+\pa{k^2-\frac{z''}{z}}v_k=0,
\ee
where the prime denotes the derivative w.r.t. conformal time, 
\be{defz}
z\equiv \frac{\sqrt{c+3\frac{\dot U^2}{\dot  h^2 U}}}{1+\frac{\dot U}{2HU}}\frac{a\dot  h}{H},
\ee
and expanding the solutions found in the long wavelength regime. In the SR approximation one finds the following expressions for the amplitude of the comoving curvature perturbations ${\mathcal R}_k\equiv v_k/z$
\be{PR}
{\mathcal P}_{\mathcal R}\simeq\frac{V_*}{24\pi^2U_* h_*^2\pa{1+3n_{U,*}^2\frac{U_*}{ h_*^2}}}\paq{\frac{\frac{ h_*^2}{2U_*}+n_{U,*}\pa{n_{V,*}-n_{U,*}/2}}{2n_{U,*}-n_{V,*}}}^2
\ee
and the scalar spectral index
\be{nsm1SR}
n_s-1\simeq \frac{-2 \pa{\de{2,*}+\de{1,*}+\ep{1_*}}-\frac{3 n_{U,*}^2\paq{2\pa{\ep{1,*}+\de{2,*}}+\pa{2 n_{n_U,*}+n_{U,*}}\de{1,*}}U_*}{ h_*^2}}{1+\frac{3 n_{U,*}^2 U_*}{ h_*^2}}.
\ee
The subscript $\!\!\!\phantom{A}_*$ indicates that the corresponding quantity must be calculated at the pivot scale, i.e. $\sim 60$ e-folds before inflation ends. These two expressions are tightly constrained by CMB observations and are thus fundamental quantities to test the viability of an inflationary model. 
Milder constraints, from present observations, are imposed on the amplitude of primordial gravitational waves ${\mathcal P}_h$ and the running of the scalar spectral index $\alpha_s$. In the SR approximation such observables have the following form
\be{Ph}
{\mathcal P}_h\simeq \frac{V_*}{12\pi^2U_*^2}
\ee
and
\be{alphas}
\alpha_s\equiv \left.\frac{\rd n_s}{\rd \ln k}\right|_{k_*}\simeq \left.\frac{\rd n_s}{\rd N}\right|_{k_*}.
\ee
In particular from (\ref{Ph}) and (\ref{PR}) one obtains the tensor to scalar ratio: 
\be{rdef}
r\equiv\frac{{\mathcal P}_h}{{\mathcal P}_{\mathcal R}}=\frac{2 h_*^2\pa{1+3n_{U,*}^2\frac{U_*}{ h_*^2}}}{U_*} \paq{\frac{2n_{U,*}-n_{V,*}}{\frac{ h_*^2}{2U_*}+n_{U,*}\pa{n_{V,*}-n_{U,*}/2}}}^{2}.
\ee
 
\section{Amplification of the spectrum}
In order for an inflationary model to generate large curvature perturbations that collapse to form PBHs after inflation ends, SR conditions must be violated at some point during inflation. Typically, such amplification can be achieved if SR transitions into a Constant Roll (CR) or Ultra Slow Roll (USR) phase \cite{USR,Chataignier:2023ago,Kamenshchik:2024kay}, during which the hierarchy of SR parameters becomes $\epsilon_{2i+1} = 2\epsilon_1$, $\epsilon_{2i} = \epsilon_2$, with $i$ a positive integer and $\ep{1}\rightarrow 0$, $\ep{2}\rightarrow {\rm const}$ in the limit $t\rightarrow \infty$. USR is a special case of CR, leading to a steeper enhancement of curvature perturbations and appears more efficient. We shall therefore restrict our analysis to the USR case.

In the Einstein Frame (EF), or when inflation is driven by a minimally coupled inflaton field $\phi$, a brief USR phase (lasting only a few e-folds) is sufficient to amplify the power spectrum of curvature perturbations {as originally suggested in \cite{USRinf}}. In this context, an inflection point in the inflaton potential $W$ (i.e., $W_{,\phi} = W_{,\phi\phi} = 0$) may be necessary, though not sufficient, to achieve the desired amplification and to construct a ``viable'' inflationary model \cite{PBHEF}. Indeed, depending on the dynamics of the preceding SR phase, the inflaton may become trapped at the inflection point, or conversely, pass through it too quickly, thereby failing to realize sufficient amplification. To achieve a more efficient enhancement, a period of quasi-USR \cite{QUUSR} may be necessary. This can be achieved by a tiny deformation in the shape of $W(\phi)$ compared to the exact USR case, which affects the deceleration of the inflaton field enhancing the spectrum. \\
{Single-field inflationary models that lead to a large amplification of perturbations at scales exiting the horizon several e-folds after the CMB ones have been widely debated in the literature in recent years. Indeed, in \cite{Quantum1}, it was argued that loop corrections—arising from the contribution of amplified short-scale modes and large couplings—may significantly alter the CMB spectrum, potentially ruling out inflationary models featuring an ultra-slow-roll (USR) phase or a similar evolution. Several papers (see \cite{quantum2} for a brief overview) have contributed to this debate, but no definitive consensus has yet been reached. \\
While general physical arguments suggest that short-scale physics should not influence long-wavelength observables, explicit calculations of quantum effects—often based on slightly different assumptions—have led to divergent and model-dependent conclusions. Without delving into the technical details, one should honestly acknowledge that a large amplification for curvature perturbations from an intermediate stage of inflation cannot currently be ruled out. However, a uniform consensus will ultimately be necessary to confidently exclude the possibility that primordial black holes (PBHs) have an inflationary origin.} 

In the case of Higgs inflation (i.e., a non-minimally coupled inflaton), the conditions for achieving a USR phase around some field value $h_0$ are given by:
\begin{equation}
2n_U(h_0) = n_V(h_0),\quad n_{n_U}(h_0) = n_{n_V}(h_0),
\label{USRJF}
\end{equation}
and one can easily verify that these conditions correspond to the existence of an inflection point in the potential $W$ of the EF-transformed model. The Eqs. (\ref{USRJF}) impose two additional constraints on the inflationary potential at field values closer to the end of inflation ($\sim 30$ e-folds before). Ensuring that the inflaton spends enough time near $h_0$ to produce sufficient amplification is a condition that must be verified by numerically solving the dynamics of the inflaton-gravity system.

If PBHs are to account for the entire DM content of the Universe, current observational constraints restrict their masses to a narrow range, approximately $\left[10^{-17}, 10^{-12}\right]M_\odot$ \cite{Carr:2020xqk,Carr:2023tpt, Ozsoy:2023ryl}. The corresponding perturbation can be estimated to exit the horizon during inflation a number of e-folds given by:
\begin{equation}
\Delta N \sim \ln \frac{a_{\mathrm{PBH}}}{a_{\mathrm{CMB}}} = \ln \alpha + \ln \frac{a_f H_f}{k_*},
\label{dNcalc}
\end{equation}
after the CMB scales exit, where $\alpha \equiv H_{\mathrm{CMB}} / H_{\mathrm{PBH}} > 1$ is the ratio of the Hubble parameter values at the CMB scales and PBH scales horizon exits, respectively. The subscript $f$ refers to the time of PBH formation, which is assumed to occur during the radiation-dominated era. Assuming standard cosmology from the Hot Big Bang to the present, one finds:
\begin{equation}
\Delta N \sim 19.6 + \ln \alpha + \frac{1}{2} \ln \frac{M_\odot}{M_{\mathrm{PBH}}}.
\label{dN}
\end{equation}
In many scenarios presented in the literature~\cite{PBHEF}, where the inflaton is minimally coupled, one has $\epsilon_1 \ll 1$ throughout the entire inflationary period, making $\ln \alpha$ negligible. In these models, for example, the perturbations associated with PBHs of mass $M_{\mathrm{PBH}} \sim 10^{-17}M_\odot$ exit the horizon approximately $39$ e-folds after the CMB pivot scale $k_* = 0.05\,\mathrm{Mpc}^{-1}$.

The seven order of magnitude of amplification is typically realised if the spectral index $n_s \sim 0.950$, i.e. is slightly smaller than the preferred value of $\sim 0.965$, but still $\sim 3$$\sigma$ within observational bounds. Let us note that, by considering a running spectral index, $\rd n_s/\rd\ln k\neq 0$, CMB observations favour smaller values of $n_s$ compared to the constant case. Then $n_s \sim 0.950$ would be $[1-2]$$\sigma$ away from its best fit. \\
{On considering the recent ACT data release \cite{ACT}, one finds that the constraints on the spectral index are looser than those from the latest Planck estimates \cite{Planck:2018jri}. However, combining the two datasets leads to tighter constraints on $n_s$, with a shift of approximately $1\sigma$ toward higher values of the spectral index ($n_s = 0.9709 \pm 0.0038$). Since inflationary models featuring a phase of ultra slow-roll (USR) typically predict a scalar spectrum that is slightly more red-tilted compared to single-field slow-roll (SR) models, higher values of $n_s$ increase the tension with inflationary scenarios that include a USR phase. Nonetheless, such models are not ruled out.}

It is worth noting that to correctly describe the full inflationary evolution—including both the SR and USR phases—and obtain a precise estimate for the amplified inflationary spectrum, one cannot rely entirely on analytical methods. The homogeneous background equations must be solved numerically, and the same applies to the Mukhanov–Sasaki (MS) equation, which must be integrated to extract the features of the primordial power spectrum. Nevertheless, analytical estimates remain useful for tuning model parameters and guiding the construction of viable models.

\section{Higgs inflation EFT}
In what follows we discuss a minimal extension of the effective action (\ref{action}) and illustrate how it is possible to construct a viable inflationary scenario featuring a SR period followed by an amplifying phase. We start from the following forms for $U$ and $V$:
\begin{equation}
U = \frac{\M^2}{2} \left(1 + u_1 \frac{h^{2}}{\M^{2}} + u_2 \frac{h^{4}}{\M^{4}}\right), \quad 
V \simeq \frac{\lambda}{4} h^4 \left(1 + v_1 \frac{h^{2}}{\M^{2}} + v_2 \frac{h^{4}}{\M^{4}} \right),
\label{UV}
\end{equation}
where the Higgs vacuum expectation value is neglected as it is irrelevant during inflation. The parameter $\lambda \sim 1/8$ is fixed at its Standard Model (SM) value. The value of $\lambda$ in the context of Higgs inflation has been extensively discussed in the literature \cite{Higgsstab}. In principle, this parameter is very sensitive to quantum corrections and may even become negative, depending in particular on the precise value of the top quark mass. If this occurs, the Higgs potential develops an instability, making inflation unviable. We leave a detailed analysis of this possibility for future studies, and for simplicity, we assume here that quantum effects have a negligible impact on its value.

Note that the addition of a quartic contribution to the non-minimal coupling $U$ necessarily implies the presence of a quartic correction ($V \propto h^8$ for large $h$) to the inflaton potential. The simultaneous presence of these contributions flattens the EF potential $W \propto V/U^2$ in the large field limit, a condition necessary for the existence of a SR phase. Higher-order contributions are neglected, as we restrict our study to the minimal extension of the original model (noting that the values of the parameters are accordingly modified).\\
{We further note that higher-derivative contributions or modifications of the standard kinetic term of the form $Z(h)(\partial h)^2$, which can indeed be generated by quantum effects, are not taken into account. The former contributions involve factors of $\pa{k/(a\M)}^n$ and are typically suppressed during inflation (and at large distances). However, if present, they could produce non-negligible effects—such as large non-Gaussianities—that are incompatible with current observations. We therefore restrict our analysis a priori to viable inflationary models and neglect such contributions. Modifications to the kinetic term of the latter type are also omitted. In fact, through a redefinition of the Higgs field, such corrections can be eliminated in the kinetic term but would then affect the functional form of $U$ and $V$. One can thus argue that the specific form chosen for $U$ and $V$ implicitly reflects a particular choice of $Z(h)$. We finally note that such corrections, being proportional to $(h/\M)^n$, are irrelevant at SM scales.}\\
{Let us further clarify that models of inflation containing higher-derivative contributions have been considered in the literature, showing that they may indeed be compatible with observations (see, for example, some of the models discussed in \cite{DeFelice2}). As a rule of thumb, this occurs if the sound speed of the generated perturbations remains close to $1$ during slow-roll (SR) inflation (see \cite{Chen} for an exhaustive review, and \cite{DeFelice} for a discussion focused on scalar-tensor theories). If this is not the case, one typically obtains large equilateral non-Gaussianities, as occurs in k-inflation, where, however, the bispectrum in the squeezed limit remain small. Indeed, a general theorem \cite{maldacena} proves that non-Gaussianities with a large squeezed limit cannot arise in single-field SR inflation, independently of the details of the inflationary action. A large bispectrum in the squeezed limit can instead be generated by a violation of SR, such as USR, as is often pointed out in inflationary models that generate large scalar perturbations, slightly modifying the predicted abundance of PBHs but not affecting CMB scales~\cite{Martin:2012pe}. However the topic is quite involved and deserves further study, which we leave for future work. We conclude by noting that in our Higgs inflation model (\ref{action}), with inflation taking place during the SR phase, one expects non-Gaussian corrections to be negligible.}

\begin{table}[t!]
  \begin{center}
    \caption{Viable Higgs inflation models}
    \label{table0}
    \begin{tabular}{|c|c|c|c|c|}
    \hline
      & $u_1$ & $u_2$ & $ x_0$ & $\epsilon$ \rule{0pt}{13pt}\\
      \hline
      \hline
     Model 1 &  $10^{-6}$ & $10^{7.5}$ & $10^{-1.938}$ & $5.98795\cdot10^{-4}$  \rule{0pt}{13pt}\\
      \hline
     Model 2 &  $-10^{-6}$ & $10^{7.51}$ & $10^{-1.941}$ & $7.02610\cdot10^{-4}$  \rule{0pt}{13pt}\\
      \hline
    \end{tabular}
  \end{center}
\end{table}

On imposing the conditions (\ref{USRJF}), one obtains the following constraints on the free parameters of the potential to ensure the existence of an USR phase:
\begin{equation}
v_1 = \frac{2\left(u_1 u_2 x_0^6 - 3 u_1 x_0^2 - 4\right)}{x_0^2\left[\left(2 u_2 + u_1^2\right) x_0^4 + 6 u_1 x_0^2 + 6\right]}, \quad 
v_2 = \frac{u_2^2 x_0^8 + 2 u_1 x_0^2 + 3}{x_0^4\left[\left(2 u_2 + u_1^2\right) x_0^4 + 6 u_1 x_0^2 + 6\right]},
\label{USRh0}
\end{equation}
where $x = h/\M$, $u_2 > 0$, and $h_0$ is the approximate field value during USR, with $x_0$ its dimensionless counterpart.

In the large field limit—where SR is expected to occur—we have $h^2/U \ll 1$ and:
\begin{equation}
n_V \sim 8 - \frac{2 v_1}{v_2 x^2}, \quad n_U \sim 4 - \frac{2 u_1}{u_2 x^2},
\label{nVnU}
\end{equation}
with:
\begin{equation}
\delta_1 \simeq \frac{\frac{v_1}{v_2} - \frac{2 u_1}{u_2}}{3 x^2}, \quad 
\delta_2 \simeq -2 \delta_1, \quad 
\epsilon_1 \simeq -2 \delta_1.
\label{SRpar}
\end{equation}
From these, we find simplified expressions for the inflationary observables:
\begin{equation}
n_s - 1 \simeq 4 \delta_{1,*}, \quad 
\mathcal{P}_\mathcal{R} \simeq \frac{2 \lambda v_2}{\left(48 \pi u_2 \delta_{1,*} \right)^2},
\label{obs}
\end{equation}
where $\lambda \simeq 1/8$, $n_s - 1 \sim -5 \cdot 10^{-2}$, and $\mathcal{P}_\mathcal{R} \simeq 2.2 \cdot 10^{-9}$. Numerical simulations show that these analytical estimates fit well the SR regime for viable models, but they fail near the USR region. However, the USR phase is characterised by a rapid variation of the SR parameters and lasts only a few e-folds. One can extrapolate the behaviour in Eq.~(\ref{SRpar}) up to $h_0$ (as is common when analysing evolution close to the end of inflation), obtaining:
\begin{equation}
\delta_1 = \frac{x_{,N}}{x} \Rightarrow 
x_0^2 \simeq x_*^2 + \frac{2}{3} \Delta N \left(\frac{v_1}{v_2} - \frac{2 u_1}{u_2} \right),
\label{SRDN}
\end{equation}
where $\Delta N \sim 35$.

Summarising: the effective action initially contains four free parameters. These are reduced to two by the conditions in Eq.~(\ref{USRh0}), though the USR point $x_0$ and pivot scale $x_*$ remain unspecified. Ultimately, only one parameter is left free after applying the observational constraints in Eqs.~(\ref{obs}) and the relation in Eq.~(\ref{SRDN}). By varying this parameter, we can identify a set of models that yield a large amplification. Moreover, we note that the simplified expressions for observational constraints (\ref{obs}) have been derived using reasonable—but not fully general—assumptions. These assumptions allow the derivation of analytically solvable equations; relaxing them would result in more complex systems, potentially intractable even numerically.

We found that the ``pure'' USR evolution produced by the polynomials $U$ and $V$ in Eq.~(\ref{UV}) is sufficient to just generate 4–5 orders of magnitude of amplification of the curvature perturbation spectrum. This would lead to the formation of an amount of PBHs insufficient to explain the entirety of present DM. A small deviation from USR—previously termed quasi-USR—is then required to keep the field near $h_0$ for a few additional e-folds, producing the necessary amplification of about 7 orders of magnitude. Such a deviation can be achieved by slightly deforming the potential around $h_0$ via the substitution $v_2 \rightarrow v_2 (1 + \epsilon)$. In the EF, this deformation creates a small bump in the otherwise flat inflection region, slowing the Higgs field locally. Without this small deformation, the inflaton would pass too rapidly through $h_0$, and no significant amplification would occur.

Let us now illustrate two of the viable models found. These models are consistent with the constraints described above (\ref{SRpar}-\ref{SRDN}), which reduce the parameter space to a single dimensionless quantity. For these models, the quasi-USR evolution occurs in the range $x_0 \in [10^{-2}, 10^{0}]$, but only values around $x_0 \sim 10^{-2}$ lead to the desired amplification. Correspondingly, only a narrow region of parameter space satisfies observational constraints and yields the entire dark matter content of the present Universe in the form of PBHs. This result is highly non-trivial, as the inflationary evolution is extremely sensitive to the parameter values. \\
In Table~\ref{table0}, we present the free parameters of two potentially viable models, which differ essentially in the sign in front of the non-minimal coupling $u_1$. For these models, the potential is given by Eq.~(\ref{UV}), with $v_1 $ and $ v_2 $ determined by Eq.~(\ref{USRh0}). The listed values are exact, and even slight variations in these parameters can lead to significantly different results. Nevertheless, we have verified that a small region of the parameter space yields (quite similar) viable models. In contrast to standard Higgs inflation, the SR phase in our scenario occurs in the regime where $ V \sim h^8 $ and $ U \sim h^4 $, requiring a very large non-minimal coupling $ u_2 $ associated with the quartic term $ h^4 $. The non-minimal coupling with the quadratic term $ h^2$, $ u_1 $, on the other hand, must be small, and it can be shown that the dynamics is largely insensitive to its precise value. We can therefore conclude that the requirement of a significant production of PBHs in the asteroid mass range (which could account for the entirety of dark matter today), while remaining consistent with CMB observations, imposes stringent constraints on the model—particularly on the parameters $v_1$, $v_2$, $ u_2$, and $x_0$—while leaving $u_1$ less constrained, provided that $u_1 \ll 1$.
\\

Table~\ref{table1} shows the corresponding inflationary observables. The values in this table are truncated approximations rather than exact numbers. The tensor-to-scalar ratio $ r $ and the running of the spectral index $ \alpha_s $ are also reported and found to be compatible with current observational constraints. Notably, the values of $ n_s $, $ r $, and $ \alpha_s $ are very similar to those obtained in other studies with a minimally coupled inflaton. By slightly modifying the duration of inflation—via different choices of model parameters—one can obtain different predictions for the PBH mass $ M_{PBH} $.

It is important to note that the higher dimension quantum operators we are considering are highly suppressed at scales well below the Planck scale (and cannot be constrained by present accelerator experiments). However, these quantum corrections could become relevant in the generation of the stochastic gravitational wave background, which may be probed by experiments like NanoGRAV \cite{NANOGrav:2023gor, DeLuca:2020agl} or the next generation of gravitational wave observatories. Therefore, in principle, these inflationary models can be tested and the calculation of the resulting GWs background is important and left for future work.\\
{Higher order monomial corrections to $U$ and $V$, similarly to the already mentioned higher-derivative contributions, are neglected. That the chosen form of the Effective Action be stable against higher order quantum correction has not been verified and certainly deserves further studies which are left for future work (in principle such corrections may have a quantum gravitational origin or may arise from SM or some beyond the SM theory). Studying such a stability is therefore highly non-trivial.\\ 
{In the context of Higgs inflation, the presence of higher-order operators has been discussed in the literature as a possible source of perturbative unitarity violation. For example, due to the non-minimal coupling $ \xi $ of the Higgs field to the Ricci scalar (see, e.g., \cite{Burgess}), the vertex $ h + h \rightarrow g $, where $ g $ denotes the graviton, leads to a violation of the unitarity bound at inflationary energy scales. Indeed, the tree-level, graviton-mediated scattering amplitude $ h + h \rightarrow h + h $ grows with energy as $ \sim (E/\Lambda)^2 $, where the cutoff scale is $ \Lambda = M/\xi \ll M $. This signals the presence of a cutoff in the EFT, which should thus be trusted only at energies $ E \ll \Lambda $. This appears to be in tension with Higgs inflation, which takes place in the ``induced gravity'' regime, $ h \gg M/\sqrt{\xi} $, i.e., at energies above this cutoff.

However, a more careful treatment of the (non-trivial) background dependence of the quantum fields (see, e.g., \cite{Bezrukov:2010jz}) leads to a different conclusion. In this approach, by properly normalizing the graviton and Higgs fluctuations in the inflationary background, one obtains a field-dependent cutoff scale, which remains larger than the typical energy scales during inflation and is thus compatible with perturbative unitarity within the EFT.

Without entering into technical details, we can outline a dimensional analysis to estimate the lowest cutoff scale in the viable Higgs inflation models considered in this article, described by the action~(\ref{action}) with~(\ref{UV}) and parameters listed in Table~\ref{table0}. On a trivial (flat space) background, we find $ \Lambda/M \sim 10^{-2} - 10^{-3} $ and $ h_* \sim 10^{-2} $, suggesting that inflation occurs near the edge of the perturbative unitarity bound. In contrast, when expanding around the homogeneous inflationary background, the cutoff scale becomes
\begin{equation}
\Lambda_h \sim \frac{\sqrt{2}(U + 2U_{,h}^2)}{2\sqrt{U}\, U_{,hh}},
\label{lambdah}
\end{equation}
which corresponds to $ \Lambda_h \sim \mathcal{O}(10)\, M $ at inflationary energies, i.e. well above the inflationary scale. This suggests that no unitarity violation occurs in this regime and that the EFT remains valid throughout inflation. The issue certainly deserves a more accurate discussion which is therefore left for future work.}\\
We stress, however, that, in the present form the model is viable and the parameter space looks very constrained by observations. The expressions (\ref{UV}) for $U$ and $V$ can originate from the quantum corrections to a scalar-theory which is supposed to describe the ``gravitational sector'' and inflation, with the Higgs dynamically generating Newton's constant. At much lower energies GR and SM are recovered.}

\begin{table}[t!]
  \begin{center}
    \caption{Inflationary Observables}
    \label{table1}
    \begin{tabular}{|c|c|c|c|c|c|}
    \hline
     &$N_{inf}$ & $M_{PBH}$ & $n_s$ & $r$ & $\alpha_s$ \rule{0pt}{13pt}\\
      \hline
      \hline
    Model 1 & $\sim 64$ & $\sim 8.4\cdot 10^{-15}M_{\odot}$ & $\sim 0.950$ & $\sim 1.1\cdot 10^{-2}$ & $\sim -1.3\cdot 10^{3}$ \rule{0pt}{13pt}\\
      \hline
    Model 2 & $\sim 63$ & $\sim 5.7\cdot 10^{-15}M_{\odot}$ & $\sim 0.950$ & $\sim 1.1\cdot 10^{-2}$ & $\sim -1.3\cdot 10^{3}$ \rule{0 pt}{13pt}\\
      \hline
    \end{tabular}
  \end{center}
\end{table}

\begin{figure}[t]
  \centering
  \includegraphics[width=0.45\textwidth]{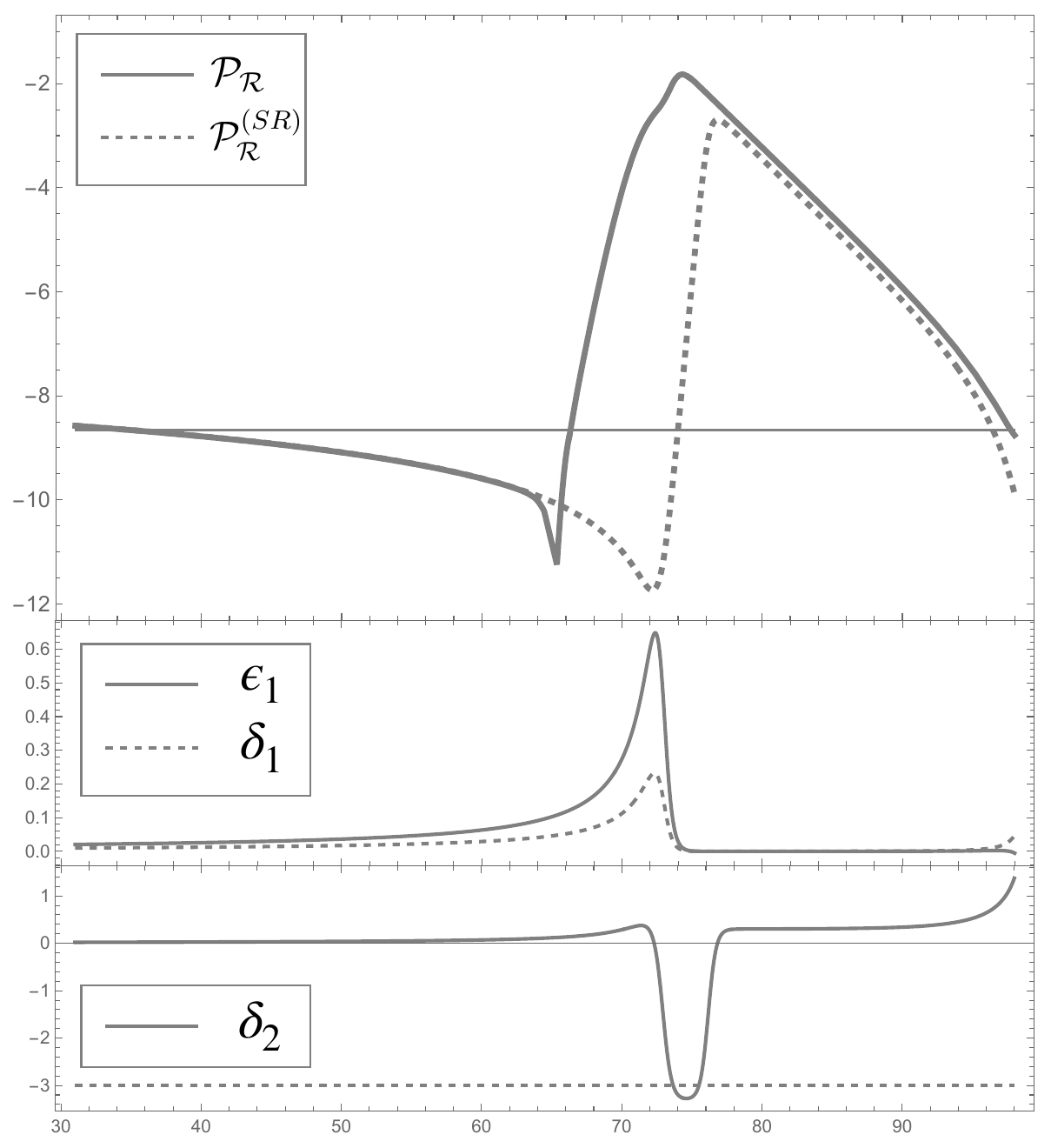}
  \includegraphics[width=0.44\textwidth]{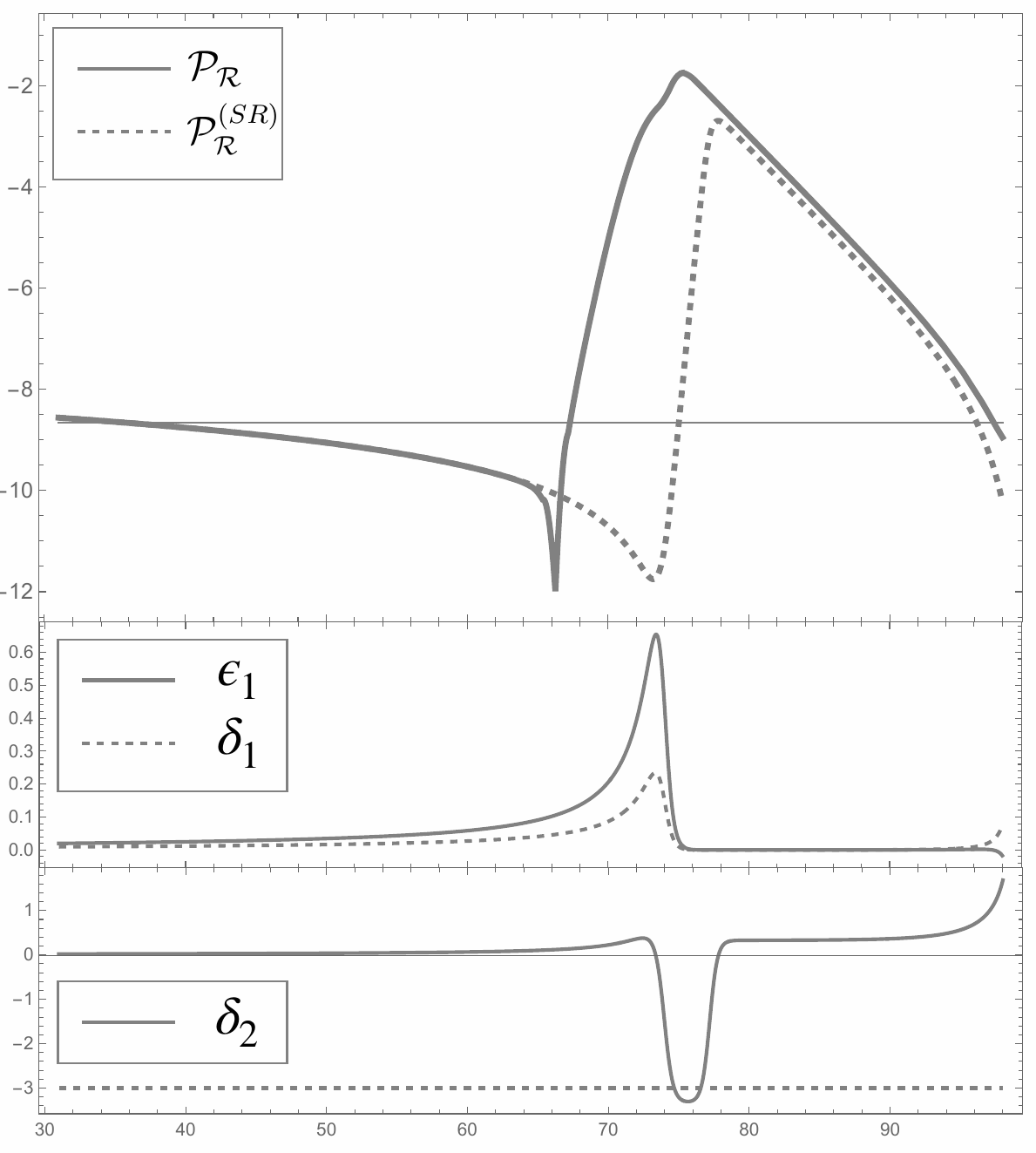}
  \caption{In the top panel, the primordial power spectrum computed numerically using the Mukhanov--Sasaki (MS) equation (solid line) is compared with the spectrum obtained using the slow-roll (SR) approximation (dashed line), for model 1 (left) and model 2 (right). The evolution of the SR parameters is shown in the lower panels. All quantities are plotted as functions of the number of e-folds $N$, which increases along the horizontal axis. The onset of inflation is identified by the point where $\mathcal{P}_{\mathcal{R}}$ intersects the horizontal line corresponding to $2.2 \cdot 10^{-9}$. Inflation ends at the far right of the plots, where the SR parameters begin to grow significantly.
}
  \label{fig}
\end{figure}

Finally, the evolution of the models and the corresponding primordial spectrum $\mathcal P_{\mathcal R}$ are illustrated in Figure~\ref{fig}. In the large panel above, the primordial spectrum computed numerically using the Mukhanov–Sasaki (MS) equation (solid line) is compared with the spectrum obtained using the SR approximation (dashed line), which fails to accurately capture the amplifying phase. The evolution of the SR parameters is shown in the lower panels. All quantities are plotted as functions of the number of e-folds $N$, which increases along the horizontal axis. For the spectrum, each value of $N$ corresponds to the time of horizon exit for the associated mode.

In particular, we observe that $\epsilon_1 \sim 5 \cdot 10^{-1}$ before the amplification begins. This is a distinctive feature of USR evolution for a non-minimally coupled inflaton, as opposed to that of a minimally coupled one, where typically the condition $\epsilon_1 \ll 1$ persists until the end of inflation. As a result, the modified relation given in Eq.~(\ref{dNcalc}) must be used in this context. Finally we observe that, during the amplification phase, the SR parameter $\delta_2$ becomes smaller than $-3$, which is a distinctive feature of quasi-USR evolution. In contrast, for pure USR or CR cases, the condition $\delta_2 \ge -3$ is always satisfied.

\section{Conclusions}
Primordial black holes have recently emerged as compelling candidates for explaining dark matter~\cite{Carr:2020xqk,Carr:2023tpt,Ozsoy:2023ryl}, as well as the unexpectedly large masses of black hole binaries detected by the LIGO/Virgo collaboration in recent years \cite{LIGO}. Additionally, PBHs could account for the features of the stochastic gravitational wave background observed by NanoGRAV and other experiments, offering valuable clues about the physics of inflation.
In fact, large primordial curvature perturbations generated during inflation can collapse into PBHs once inflation ends.

Over the past decade, several inflationary models and mechanisms have been studied to explain the amplification of inflationary perturbations. Many of these models employ a phase of ultra slow-roll (USR) to achieve the desired amplification. For the case of minimally coupled inflation, such a phase can be readily obtained through the presence of an inflection point---or quasi-inflection point---in the inflaton potential, depending on its slope. This suggests that the structure of the inflaton potential at scales well below those probed by the CMB may be much richer and could reveal new physics.

Inflation driven by a non-minimally coupled inflaton has been investigated in several articles  \cite{infNMC}. In particular, Higgs inflation has attracted considerable attention (see for example \cite{hinf}), as it fits current observational data well and requires no new physics beyond the Standard Model. The presence of a non-minimal coupling can, in fact, be naturally justified as a consequence of quantum corrections. Nonetheless, to our knowledge, only a few papers \cite{pbhNM} have explored the possibility of generating large curvature perturbations in the presence of a non-minimally coupled inflaton.

In this manuscript, we have studied Higgs inflation with a particular focus on the generation of large inflationary perturbations that may subsequently collapse into PBHs at the end of inflation. {This assuming that loop corrections are small, a point which has been much debated in the literature for a minimally coupled inflaton [18,19],} USR may still be the mechanism responsible for the amplification in the non-minimally coupled case. However, as shown in~\cite{Chataignier:2023ago,Kamenshchik:2024kay}, in this context USR is not associated with the presence of an inflection point and instead exhibits dynamical features distinct from the minimally coupled case.

To fit the CMB data while still achieving amplification later during inflation, the Higgs inflation mechanism must be extended beyond the standard scenario. In this article, we showed that a viable setup can be obtained by enlarging the space of free parameters. This is done by considering the effective action of the inflaton-gravity system and including further quantum corrections in the form of a tower of higher-order operators. While these operators are suppressed at Standard Model energy scales, they play a crucial role during inflation. Moreover, their presence could become relevant in the generation of the stochastic gravitational wave background, which may be probed by experiments such as NanoGRAV or by the next generation of gravitational wave experiments.\\

By then considering a minimal extension of the original Higgs inflation model~\cite{Bezrukov:2007ep}, we investigated the existence of parameter configurations that enable successful inflation and also account for the entirety of observed dark matter in terms of PBHs. Despite the introduction of four new free parameters to the original model, achieving this goal proves highly non-trivial. Nevertheless, we demonstrated that a narrow region in parameter space satisfies these requirements, and we presented a couple of explicit models that realise this scenario.

Unlike the setup in~\cite{Bezrukov:2007ep}, in these models the slow-roll phase must occur at energy scales where \( V \sim h^8 \) and \( U \sim h^4 \), requiring a very large non-minimal coupling \( u_2 \) with the quartic term \( h^4 \). In contrast to~\cite{Bezrukov:2007ep}, where only the CMB data were fitted, our model involved a small non-minimal coupling \( u_1 \) with the quadratic term \( h^2 \). Further, due to the polynomial form of the functions \( U \) and \( V \), USR alone is insufficient to produce the required seven orders of magnitude amplification in the power spectrum. A small deformation (quasi-USR) was necessary to achieve the desired enhancement.\\



\begin{thebibliography}{99}
\bibitem{ATLAS2012}
ATLAS Collaboration,
\textit{Phys. Lett. B} \textbf{716}, 1–29 (2012),
\bibitem{CMS2012}
CMS Collaboration,
\textit{Phys. Lett. B} \textbf{716}, 30–61 (2012)

\bibitem{Bezrukov:2007ep}
F.~L.~Bezrukov and M.~Shaposhnikov,
Phys. Lett. B \textbf{659} (2008) 703-706

\bibitem{Planck:2018jri}
Y.~Akrami \textit{et al.} [Planck],
Astron. Astrophys. \textbf{641} (2020), A10


\bibitem{Sakharov:1967pk}
A.~D.~Sakharov,
Dokl. Akad. Nauk Ser. Fiz. \textbf{177} (1967), 70-71

\bibitem{Zee:1978wi}
A.~Zee,
Phys. Rev. Lett. \textbf{42} (1979), 417
doi:10.1103/PhysRevLett.42.417


\bibitem{Starobinsky:1980te}
A.~A.~Starobinsky,
Phys. Lett. B \textbf{91} (1980), 99-102

\bibitem{Novikov:1967tw}
{Y.~B.~Zel'dovich and I.~D.~Novikov,
Sov. Astron. \textbf{10} (1967) 602}

\bibitem{Hawking:1971ei}
S.~Hawking,
Mon. Not. Roy. Astron. Soc. \textbf{152} (1971) 75


\bibitem{Cerioni:2009kn}
A.~Cerioni, F.~Finelli, A.~Tronconi and G.~Venturi,
Phys. Lett. B \textbf{681} (2009) 383--386

A.~Cerioni, F.~Finelli, A.~Tronconi and G.~Venturi,
Phys. Rev. D \textbf{81} (2010), 123505

\bibitem{Hwang:1995bv}
J.~c.~Hwang,
Phys. Rev. D \textbf{53} (1996), 762-765

\bibitem{USR}

J.~Yokoyama,
Phys. Rev. D \textbf{59} (1999) 107303

R.~Saito, J.~Yokoyama and R.~Nagata,
JCAP \textbf{06} (2008), 024

K.~Dimopoulos,
Phys. Lett. B \textbf{775} (2017), 262-265

W.~H.~Kinney,
Phys. Rev. D \textbf{72} (2005), 023515

S.~M.~Leach, M.~Sasaki, D.~Wands and A.~R.~Liddle,
Phys. Rev. D \textbf{64} (2001), 023512

H.~Motohashi, A.~A.~Starobinsky and J.~Yokoyama,
JCAP \textbf{09} (2015) 018


Z.~Yi and Y.~Gong,
JCAP \textbf{03} (2018) 052



\bibitem{Chataignier:2023ago}
L.~Chataignier, A.~Y.~Kamenshchik, A.~Tronconi and G.~Venturi,
Phys. Rev. D \textbf{107} (2023) 083506

\bibitem{Kamenshchik:2024kay}
A.~Y.~Kamenshchik, E.~O.~Pozdeeva, A.~Tribolet, A.~Tronconi, G.~Venturi and S.~Y.~Vernov,
Phys. Rev. D \textbf{110} (2024) no.10, 104011

\bibitem{USRinf}
 P. Ivanov,  P. Naselsky, I. Novikov,
\emph{Phys. Rev. D}, 1994, \textbf{50}, p.~7173--7178

\bibitem{PBHEF}

J.~Yokoyama,
Phys. Rev. D \textbf{58} (1998), 083510


C.~Germani and T.~Prokopec,
Phys. Dark Univ. \textbf{18} (2017), 6-10

M.~Cicoli, V.~A.~Diaz and F.~G.~Pedro,
JCAP \textbf{06} (2018), 034

A.~Karam, N.~Koivunen, E.~Tomberg, V.~Vaskonen and H.~Veerm\"ae,
JCAP \textbf{03} (2023) 013




H.~Motohashi, S.~Mukohyama and M.~Oliosi,
JCAP \textbf{03} (2020), 002


A.~Y.~Kamenshchik, A.~Tronconi and G.~Venturi,
JCAP \textbf{01} (2022) no.01, 051

 S.V. Ketov,
\emph{Universe}, 2021, \textbf{7}, no.5, 115

A.~Y.~Kamenshchik, A.~Tronconi, T.~Vardanyan and G.~Venturi,
Phys. Lett. B \textbf{791} (2019), 201-205


\bibitem{QUUSR}
G.~Ballesteros, J.~Rey, M.~Taoso and A.~Urbano,
JCAP \textbf{08} (2020), 043

J.~Garcia-Bellido and E.~Ruiz Morales,
Phys. Dark Univ. \textbf{18} (2017), 47-54

\bibitem{Quantum1}
J.~Kristiano and J.~Yokoyama,
Phys. Rev. Lett. \textbf{132} (2024) 221003

\bibitem{quantum2}
J.~Fumagalli,
JHEP \textbf{01} (2025), 108

G.~Ballesteros and J.~G.~Egea,
JCAP \textbf{07} (2024), 052

Y.~Tada, T.~Terada and J.~Tokuda,
JHEP \textbf{01} (2024), 105

G.~Franciolini, A.~Iovino, Junior., M.~Taoso and A.~Urbano,
Phys. Rev. D \textbf{109} (2024) no.12, 123550

H.~Firouzjahi,
Phys. Rev. D \textbf{109} (2024) no.4, 043514

K.~Inomata,
Phys. Rev. Lett. \textbf{133} (2024) no.14, 141001

A.~Riotto,
[arXiv:2301.00599 [astro-ph.CO]].





\bibitem{Carr:2020xqk}
B.~Carr, F.~Kuhnel,
\emph{Ann. Rev. Nucl. Part. Sci.}, 2020, \textbf{70},  p. 355--394

\bibitem{Carr:2023tpt}
B.~Carr, S.~Clesse, J.~Garcia-Bellido, M.~Hawkins and F.~Kuhnel,
Phys. Rept. \textbf{1054} (2024) 1--68


\bibitem{Ozsoy:2023ryl}  
O. \"Ozsoy, G. Tasinato, ``Inflation and Primordial Black Holes,''
\emph{Universe}, 2023, \textbf{9}, no.5, 203



\bibitem{ACT}
E.~Calabrese \textit{et al.} [ACT],
[arXiv:2503.14454 [astro-ph.CO]]


\bibitem{Higgsstab}
G.~Isidori, G.~Ridolfi, and A.~Strumia, 
\textit{Nucl. Phys. B}, vol.~609, pp.~387--409, 2001

V.~Branchina and E.~Messina,
Phys. Rev. Lett. \textbf{111} (2013), 241801

F.~Bezrukov, J.~Rubio and M.~Shaposhnikov,
Phys. Rev. D \textbf{92} (2015) no.8, 083512

A.~Kobakhidze and A.~Spencer-Smith, 
\textit{Phys. Lett. B}, vol.~722, pp.~130--134, 2013

P.~Burda, R.~Gregory, and I.~G.~Moss, 
\textit{Phys. Rev. Lett.}, vol.~115, p.~071303, 2015

J.~R.~Espinosa, D.~Racco, and A.~Riotto, 
\textit{Phys. Rev. Lett.}, vol.~120, p.~121301, 2018

J.~R.~Espinosa \textit{et al.}, 
\textit{JHEP}, vol.~2015, no.~9, p.~174, 2015.

D.~G.~Figueroa, A.~Rajantie, and F.~Torrenti, 
\textit{JCAP}, vol.~2017, no.~10, p.~071, 2017

J.~R.~Espinosa, D.~Racco, and A.~Riotto, 
\textit{Eur. Phys. J. C}, vol.~78, p.~806, 2018

T.~Markkanen, A.~Rajantie, and S.~Stopyra, 
``Cosmological aspects of Higgs vacuum metastability,'' 
\textit{Front. Astron. Space Sci.}, vol.~5, p.~40, 2018


\bibitem{DeFelice2}
A.~De Felice, S.~Tsujikawa, J.~Elliston and R.~Tavakol,
JCAP \textbf{08} (2011), 021

A.~De Felice and S.~Tsujikawa,
JCAP \textbf{03} (2013), 030

\bibitem{Chen}
X.~Chen,
Adv. Astron. \textbf{2010} (2010), 638979

\bibitem{DeFelice}
A.~De Felice and S.~Tsujikawa,
Phys. Rev. D \textbf{84} (2011), 083504

\bibitem{maldacena}
J.~M.~Maldacena,
JHEP \textbf{05} (2003), 013

P.~Creminelli and M.~Zaldarriaga,
JCAP \textbf{10} (2004), 006

\bibitem{Martin:2012pe}
J.~Martin, H.~Motohashi and T.~Suyama,
Phys. Rev. D \textbf{87} (2013) no.2, 023514


\bibitem{NANOGrav:2023gor}
G.~Agazie \textit{et al.} [NANOGrav],
Astrophys. J. Lett. \textbf{951} (2023) no.1, L8

\bibitem{DeLuca:2020agl}
V.~De Luca, G.~Franciolini and A.~Riotto,
Phys. Rev. Lett. \textbf{126} (2021) no.4, 041303
doi:10.1103/PhysRevLett.126.041303

\bibitem{Burgess}
C.~P.~Burgess, H.~M.~Lee and M.~Trott,
JHEP \textbf{07} (2010), 007

\bibitem{Bezrukov:2010jz}
F.~Bezrukov, A.~Magnin, M.~Shaposhnikov and S.~Sibiryakov,
JHEP \textbf{01} (2011), 016




\bibitem{LIGO}
B.~P.~Abbott \textit{et al.} [LIGO Scientific and Virgo],
Phys. Rev. Lett. \textbf{116} (2016) no.6, 061102




\bibitem{infNMC}

B.~L.~Spokoiny,
Phys. Lett. B \textbf{147} (1984) 39--43

F.~S.~Accetta, D.~J.~Zoller and M.~S.~Turner,
Phys. Rev. D \textbf{31} (1985) 3046

T.~Futamase and K.~i.~Maeda,
Phys. Rev. D \textbf{39} (1989) 399--404

A.~Tronconi,
JCAP \textbf{07} (2017), 015

\bibitem{hinf}

J.~M.~Ezquiaga, J.~Garcia-Bellido and E.~Ruiz Morales,
Phys. Lett. B \textbf{776} (2018), 345-349


J.~L.~F.~Barbon and J.~R.~Espinosa,
Phys.\ Rev.\ D \textbf{79}, 081302 (2009).

F.~Bezrukov, A.~Magnin, M.~Shaposhnikov and S.~Sibiryakov,
JHEP \textbf{01} (2011), 016

F.~Bezrukov and M.~Shaposhnikov,
Phys. Lett. B \textbf{734} (2014), 249-254

K.~Kamada, T.~Kobayashi, T.~Takahashi, M.~Yamaguchi, and J.~Yokoyama,
Phys.\ Rev.\ D \textbf{86}, 023504 (2012)

A.~S.~Koshelev and A.~Tokareva,
Phys.\ Rev.\ D \textbf{102}, 123518 (2020)

A.~O.~Barvinsky, A.~Y.~Kamenshchik and A.~A.~Starobinsky,
JCAP \textbf{11} (2008) 021

A.~O.~Barvinsky, A.~Y.~Kamenshchik, C.~Kiefer, A.~A.~Starobinsky and C.~Steinwachs,
JCAP \textbf{12} (2009), 003

J.~Rubio,
Front. Astron. Space Sci. \textbf{5} (2019), 50

\bibitem{pbhNM}

F.~Bezrukov, M.~Pauly and J.~Rubio,
JCAP \textbf{02} (2018), 040

S.~Rasanen and E.~Tomberg,
JCAP \textbf{01} (2019), 038

S.~R.~Geller, W.~Qin, E.~McDonough, and D.~I.~Kaiser,
\textit{Phys. Rev. D}, vol.~106, p.~063535, 2022


E.~O.~Pozdeeva and S.~Y.~Vernov,
arXiv:2407.00999

Z.~Yi,
JCAP \textbf{03} (2023), 048







\end{thebibliography}
\end{document}